\documentclass[hidelinks,twocolumn,showpacs,aps,prl,superscriptaddress,floatfix]{revtex4}
\usepackage{graphicx}
\usepackage{verbatim}
\usepackage{dcolumn}
\usepackage{amsmath}
\usepackage{epsfig}
\usepackage{subfigure}
\usepackage{url}

\usepackage{breakurl}
\usepackage[breaklinks]{hyperref}

\newcommand{\BaBarYear}      {20}
\newcommand{\BaBarNumber}    {003}
\newcommand{\BaBarType}      {PUB}  
\newcommand{\SLACPubNumber}  {17526}

\input babarsym.tex

\def\epem{e^+e^-}
\def\mpmm{\mu^+\mu^-}
\def\tptm{\tau^+\tau^-}

\def\pipm{\pi^+\pi^-}
\def\pL{\phi_L}
\def\mpL{m_{\phi_L}}
\def\MeV{\mev}
\def\GeV{\gev}

\def\figurebox#1#2#3{%
    \def\arg{#3}%
    \ifx\arg\empty
    {\hfill\vbox{\hsize#2\hrule\hbox to #2{\vrule\hfill\vbox to #1{\hsize#2\vfill}\vrule}\hrule}\hfill}%
    \else
    {\hfill\epsfbox{#3}\hfill}%
    \fi}

\begin{document}

\pagestyle{plain}

\begin{flushleft}
\babar-\BaBarType-\BaBarYear/\BaBarNumber \\
SLAC-PUB-\SLACPubNumber \\
arXiv:2005.01885 [hep-ex] \\
\end{flushleft}

\title{{\large \bf Search for a Dark Leptophilic Scalar in $\epem$ Collisions}}

\author{J.~P.~Lees}
\author{V.~Poireau}
\author{V.~Tisserand}
\affiliation{Laboratoire d'Annecy-le-Vieux de Physique des Particules (LAPP), Universit\'e de Savoie, CNRS/IN2P3,  F-74941 Annecy-Le-Vieux, France}
\author{E.~Grauges}
\affiliation{Universitat de Barcelona, Facultat de Fisica, Departament ECM, E-08028 Barcelona, Spain }
\author{A.~Palano}
\affiliation{INFN Sezione di Bari and Dipartimento di Fisica, Universit\`a di Bari, I-70126 Bari, Italy }
\author{G.~Eigen}
\affiliation{University of Bergen, Institute of Physics, N-5007 Bergen, Norway }
\author{D.~N.~Brown}
\author{Yu.~G.~Kolomensky}
\affiliation{Lawrence Berkeley National Laboratory and University of California, Berkeley, California 94720, USA }
\author{M.~Fritsch}
\author{H.~Koch}
\author{T.~Schroeder}
\affiliation{Ruhr Universit\"at Bochum, Institut f\"ur Experimentalphysik 1, D-44780 Bochum, Germany }
% ...
\collaboration{The \babar\ Collaboration}
\noaffiliation

\begin{abstract}
Many scenarios of physics beyond the Standard Model predict the existence of new gauge singlets, which might 
be substantially lighter than the weak scale. The experimental constraints on additional scalars with masses 
in the \MeV to \GeV range could be significantly weakened if they interact predominantly with leptons 
rather than quarks. At an $\epem$ collider, such a leptophilic scalar ($\pL$) would be produced predominantly 
through radiation from a $\tau$ lepton. We report herein a search for $\epem \rightarrow \tau^+\tau^-\pL, 
\pL \rightarrow \ell^+\ell^-$ ($\ell=e,\mu$) using data collected by the \babar\ experiment at SLAC. No significant 
signal is observed, and we set limits on the $\pL$ coupling to leptons in the range $0.04\GeV < \mpL < 7.0 \GeV$. 
These bounds significantly improve upon the current constraints, excluding almost entirely the parameter 
space favored by the observed discrepancy in the muon anomalous magnetic moment below $4 \GeV$ at 90\% confidence level. 
\end{abstract}

\maketitle

\setcounter{footnote}{0}

Many theories beyond the Standard Model (SM) predict the existence of additional scalars, and discovering or constraining 
their existence might shed light on the physics of electroweak symmetry breaking and the Higgs sector (e.g., see 
Ref.~\cite{Branco:2011iw}). Some of these particles may be substantially lighter than the weak scale, notably in 
the Next-to-Minimal Supersymmetric Standard Model~\cite{Dobrescu:2000yn}, but also in more generic singlet-extended 
sectors~\cite{Chen:2015vqy, Batell:2016ove}. In the $\MeV-\GeV$ range, new scalars could mediate 
interactions between the SM and dark matter, as well as account for the discrepancy in the observed value of the muon 
anomalous magnetic dipole moment~\cite{Leveille:1977rc,Bennett:2006fi,Tanabashi:2018oca}.

The possible coupling of a new scalar $\pL$ to SM particles is constrained by SM gauge invariance. In the simplest case, the mixing 
between the scalar and the SM Higgs boson gives rise to couplings proportional to SM fermion masses. Because the new scalar
couples predominantly to heavy-flavor quarks, this minimal scenario is strongly constrained by searches for rare flavor-changing 
neutral current decays of mesons, such as $B\rightarrow K\phi$ and $K\rightarrow \pi\phi$~\cite{Beacham:2019nyx}. However, these 
bounds are evaded if the coupling of the scalar to quarks is suppressed and the scalar interacts preferentially with heavy-flavor 
leptons~\cite{Chen:2015vqy,Batell:2016ove,Liu:2016qwd,Liu:2020qgx}. We refer to such a particle as a leptophilic scalar, $\pL$. 
Its interaction Lagrangian with leptons can be described by:
$$\mathcal{L} = -\xi\sum_{\ell = e,\mu,\tau} \frac{m_\ell}{v} \bar\ell\,\pL \ell,$$
where $\xi$ denotes the flavor-independent coupling strength to leptons and $v = 246 \GeV$ is the SM Higgs vacuum expectation 
value~\cite{naturalUnits}. Model independent constraints relying exclusively on the coupling to leptons are derived from 
a \babar\ search for a muonic dark force~\cite{TheBABAR:2016rlg} and beam dump experiments~\cite{Bjorken:1988as,Davier:1989wz}. 
A large fraction of the parameter space, including the region favored by the measurement of the 
muon anomalous magnetic moment, is still unexplored~\cite{Batell:2016ove,Liu:2016qwd,Liu:2020qgx}. Examples 
of model dependent bounds from $B$ and $h$ decays for a specific UV-completion of the theory can be found in Ref.~\cite{Batell:2016ove}.
 
The large sample of $\tau^+\tau^-$ pairs collected by \babar\ offers a clean environment to study model independent 
$\pL$ production via final-state radiation in $\epem \rightarrow \tau^+\tau^-\pL$. The mass-proportionality 
of the coupling, in particular the feeble interaction with electrons, dictates the experimental signature. For 
$2m_e < \mpL < 2m_\mu$, the scalar decays predominantly into electrons, leading to displaced vertices for sufficiently 
small values of the coupling. Prompt decays into a pair of muons (taus) dominate when $2m_\mu \le \mpL < 2m_\tau$ 
($2m_\tau < \mpL$). 

We report herein the first search for a leptophilic scalar in the reaction $\epem \rightarrow \tau^+\tau^-\pL, \pL 
\rightarrow \ell^+\ell^-$ ($\ell=e,\mu$) for $0.04 \GeV < \mpL < 7.0 \GeV$. The cross section for $\mpL < 2 m_\mu$ 
is measured separately for $\pL$ lifetimes corresponding to $c\tau_{\phi_L}$ values of 0, 1, 10 and 100\,mm. Above 
the dimuon threshold, we determine the cross section for prompt $\pL \rightarrow \mpmm$ decays. In all cases, the 
$\pL$ width is much smaller than the detector resolution, and the signal can be identified as a narrow peak in the
dilepton invariant mass.

The search is based on 514\invfb of data collected at the $\Y2S,\Y3S,\Y4S$ resonances and their vicinities~\cite{Lees:2013rw} 
by the \babar\ experiment at the SLAC PEP-II $\epem$ collider. The \babar\ detector is described in detail 
elsewhere~\cite{Bib:Babar,TheBABAR:2013jta}. A sample corresponding to about 5\% of the data, called the optimization 
sample, is used to optimize the search strategy and is subsequently discarded. The remaining data are examined 
only once the analysis procedure has been finalized.

Signal Monte Carlo (MC) samples with prompt decays are simulated for 36 different $\pL$ mass hypotheses by 
the \textsc{MadGraph} event generator~\cite{Alwall:2014hca} and showered using \textsc{Pythia} 8~\cite{Sjostrand:2014zea}, 
including final-state radiation. For $\mpL < 0.3 \GeV$, events with $c\tau_{\phi_L}$ values up to 300\,mm are also 
generated. We simulate the following reactions to study the background: 
$\epem \rightarrow \epem (\gamma)$ (\textsc{BHWIDE}~\cite{Jadach:1995nk}), $\epem \rightarrow \mpmm (\gamma)$ and 
$\epem \rightarrow \tau^+ \tau^- (\gamma)$ (\textsc{KK} with the \textsc{TAUOLA} library~\cite{Jadach:2000ir,Jadach:1993hs}),  
$\epem \rightarrow q\overline{q}$ with $q=u,d,s,c$ (\textsc{JETSET}~\cite{Sjostrand:1993yb}), and $\epem \rightarrow B\bar{B}$ 
and generic $\epem \rightarrow \Upsilon(2S,3S)$ decays (\textsc{EvtGen}~\cite{Lange:2001uf}). The resonance production 
$\epem \rightarrow \gamma \psitwos, \psitwos \rightarrow \pipm\jpsi, \jpsi \rightarrow \mpmm$ is simulated with \textsc{EvtGen} 
using a structure function technique~\cite{Bib::Struct1,Bib::Struct2}. The detector acceptance and reconstruction 
efficiencies are estimated with a simulation based on \textsc{GEANT4}~\cite{Agostinelli:2002hh}.

We select events containing exactly four charged tracks with zero net charge, focusing on $\tau$ lepton decays 
to single tracks and any number of neutral particles. The $\pL \rightarrow \ell^+ \ell^-$ candidates are formed 
by combining two opposite-sign tracks identified as an electron or muon pair by particle identification (PID)
algorithms~\cite{Bib:Babar,TheBABAR:2016rlg}. We do not attempt to select a single $\pL$ candidate per event, but 
simply consider all possible combinations. Radiative Bhabha and dimuon events in which the photon converts to 
an $\epem$ pair are suppressed by rejecting events with a total visible mass greater than $9\GeV$. We further veto 
$\epem \rightarrow \epem \epem$ events by requiring the cosine of the angle between the momentum of the $\pL$ 
candidate and that of the nearest track to be less than 0.98, the missing momentum against all tracks and 
neutral particles to be greater than $300 \MeV$, and that there be three or less tracks identified as electrons. 
We perform a kinematic fit to the selected $\pL$ candidates, constraining the two tracks to originate from the same 
point in space. The dimuon production vertex is required to be compatible with the beam interaction region, while we 
only constrain the momentum vector of the $\epem$ pair to point back to the beam interaction region since the 
dielectron vertex can be substantially displaced. We select dielectron (dimuon) combinations with a value of 
the $\chi^2$ per degree of freedom of the fit, $\chi^2/n.d.f.$, less than 3 (12).

A multivariate selection based on boosted decision trees (BDT) further improves the signal purity~\cite{BDT}. The BDTs 
include variables capturing the typical $\tau$ and $\pL$ decay characteristics: a well-reconstructed $\ell^+\ell^-$ vertex, 
either prompt or displaced; missing energy and momentum due to neutrino emission; relatively large track momenta; low 
neutral particle multiplicity; and two or more tracks identified as electrons or muons. A few variables 
are also targeted at specific backgrounds, such as $\psitwos \rightarrow \pipm \jpsi, \jpsi \rightarrow \mpmm$ production 
in initial-state radiation (ISR) events. The $\pL$ mass is specifically excluded to limit potential bias in the 
classifier. A full description of these variables can be found in the Supplemental Material~\cite{SPM}. We train a separate 
BDT for each of the different final states and $c\tau_{\phi_L}$ values with signal events modeled using a flat $\mpL$ distribution 
and background events modeled using the optimization sample data. 

\begin{figure}[t]
\begin{center}
  \includegraphics[width=0.48\textwidth]{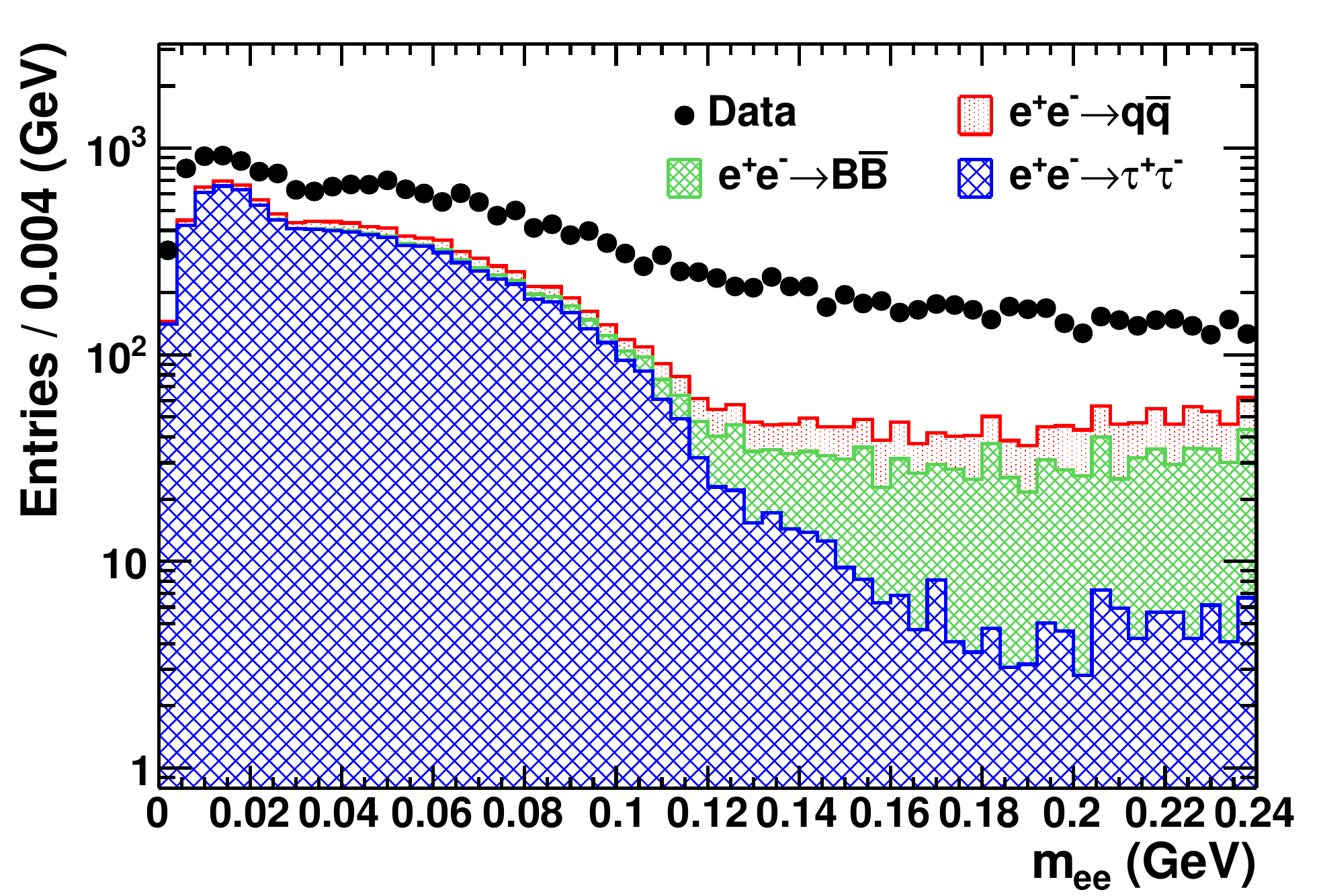}
  \includegraphics[width=0.48\textwidth]{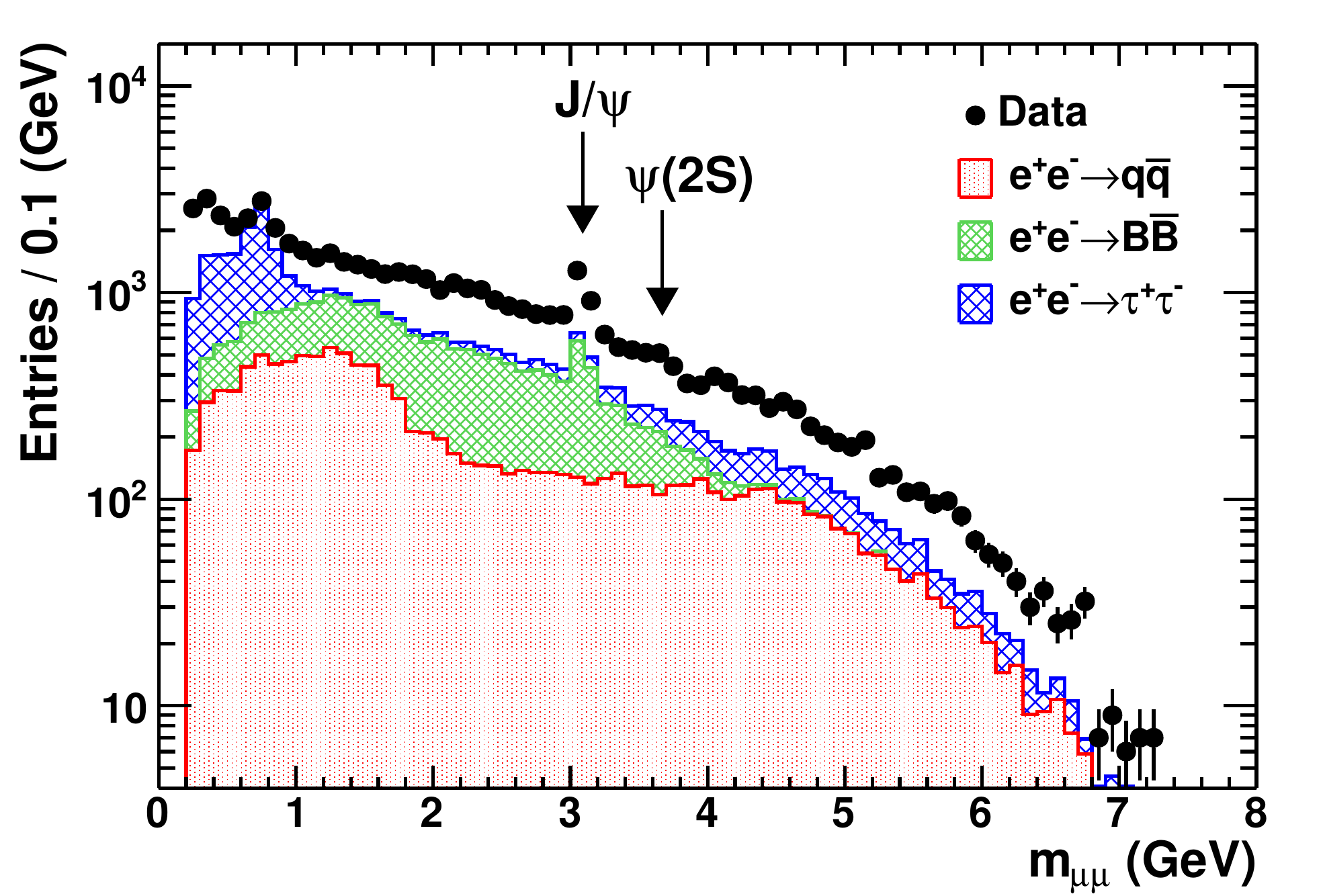}
\end{center}
\caption
{The distribution of (top) the dielectron invariant mass and (bottom) the dimuon invariant mass for prompt decays, 
together with simulated predictions for the indicated processes normalized to the integrated luminosity of the data 
(stacked histograms).}
\label{Fig1}
\end{figure}

\begin{figure}[htb]
\begin{center}
  \includegraphics[width=0.45\textwidth]{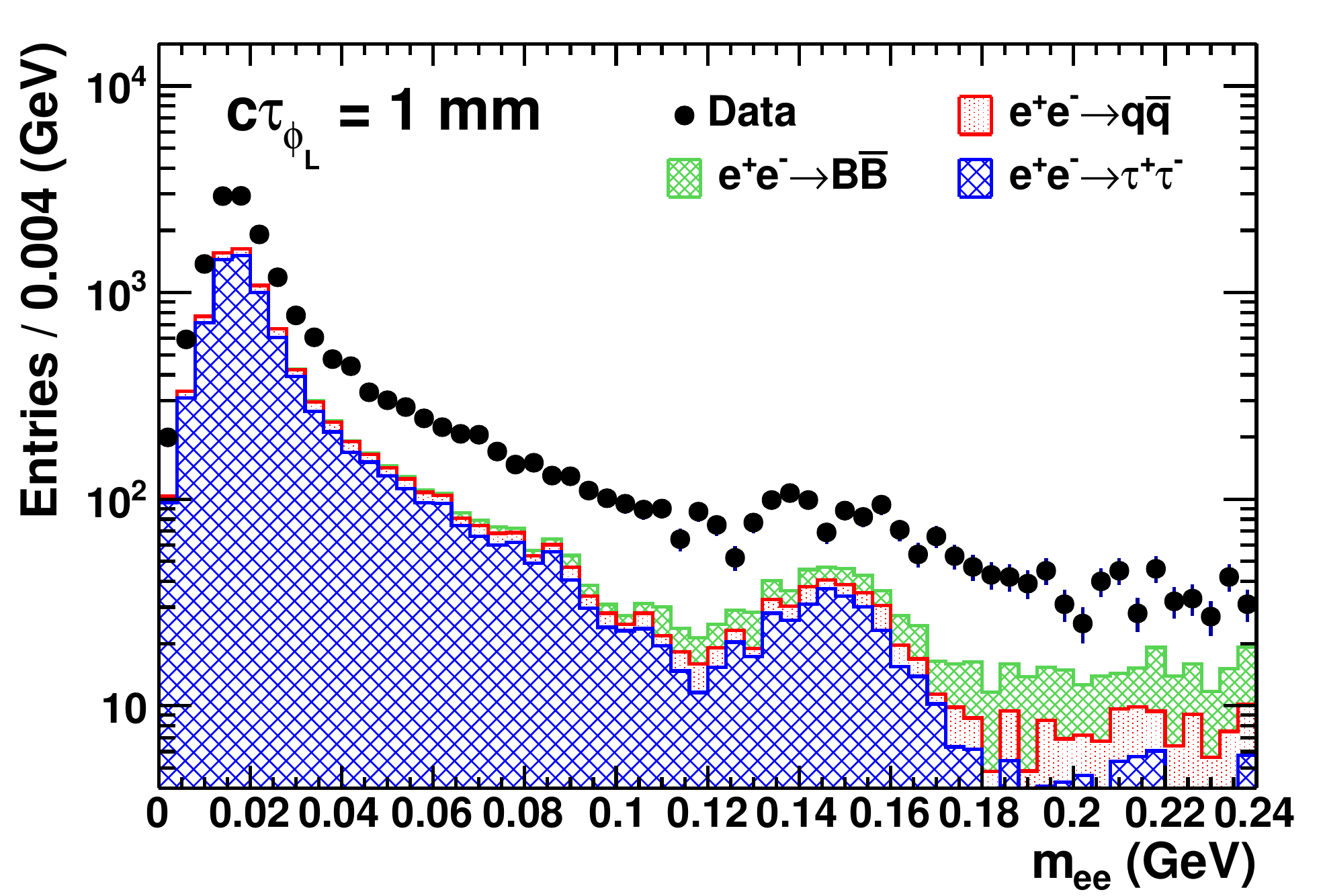}
  \includegraphics[width=0.45\textwidth]{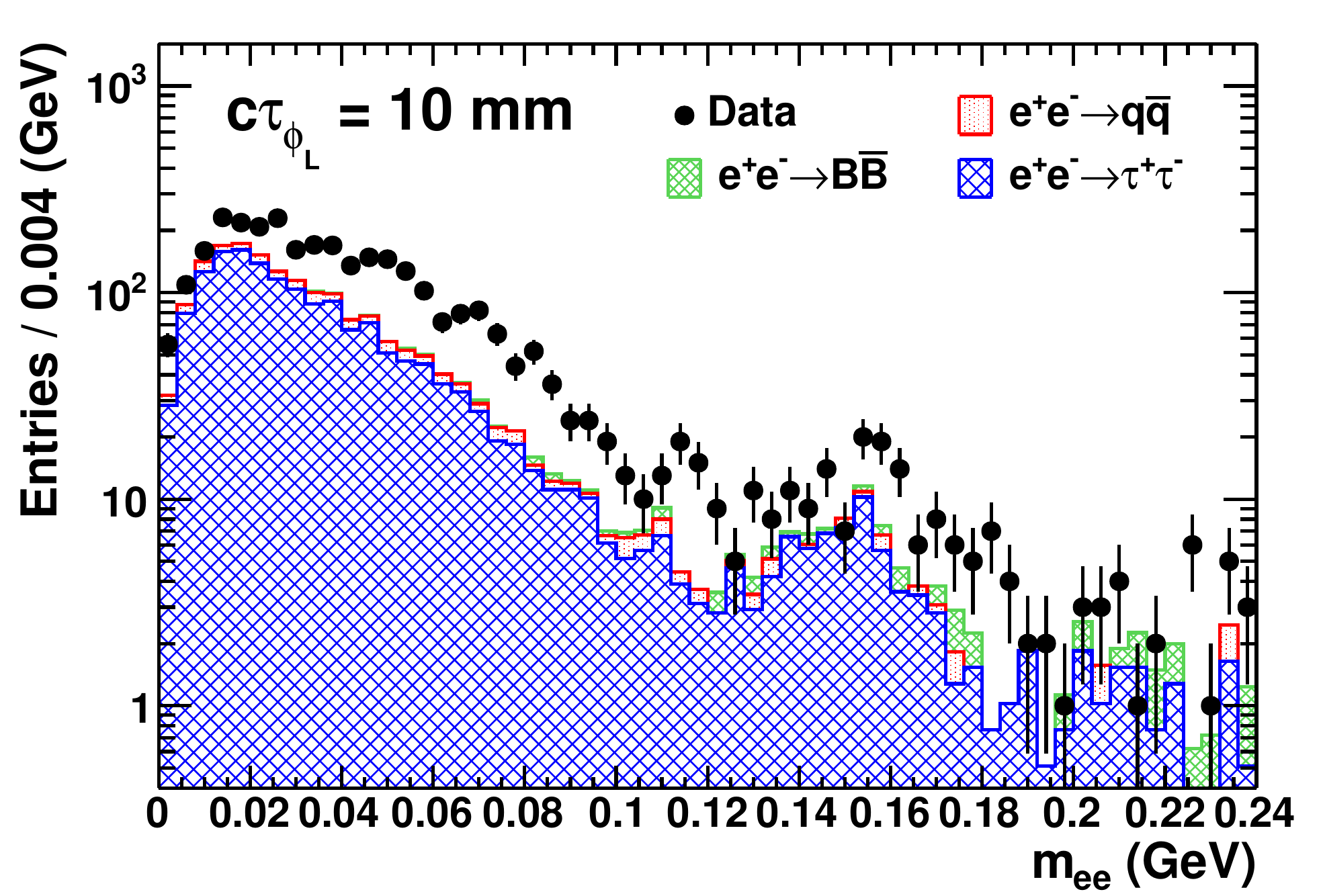}\\
  \includegraphics[width=0.45\textwidth]{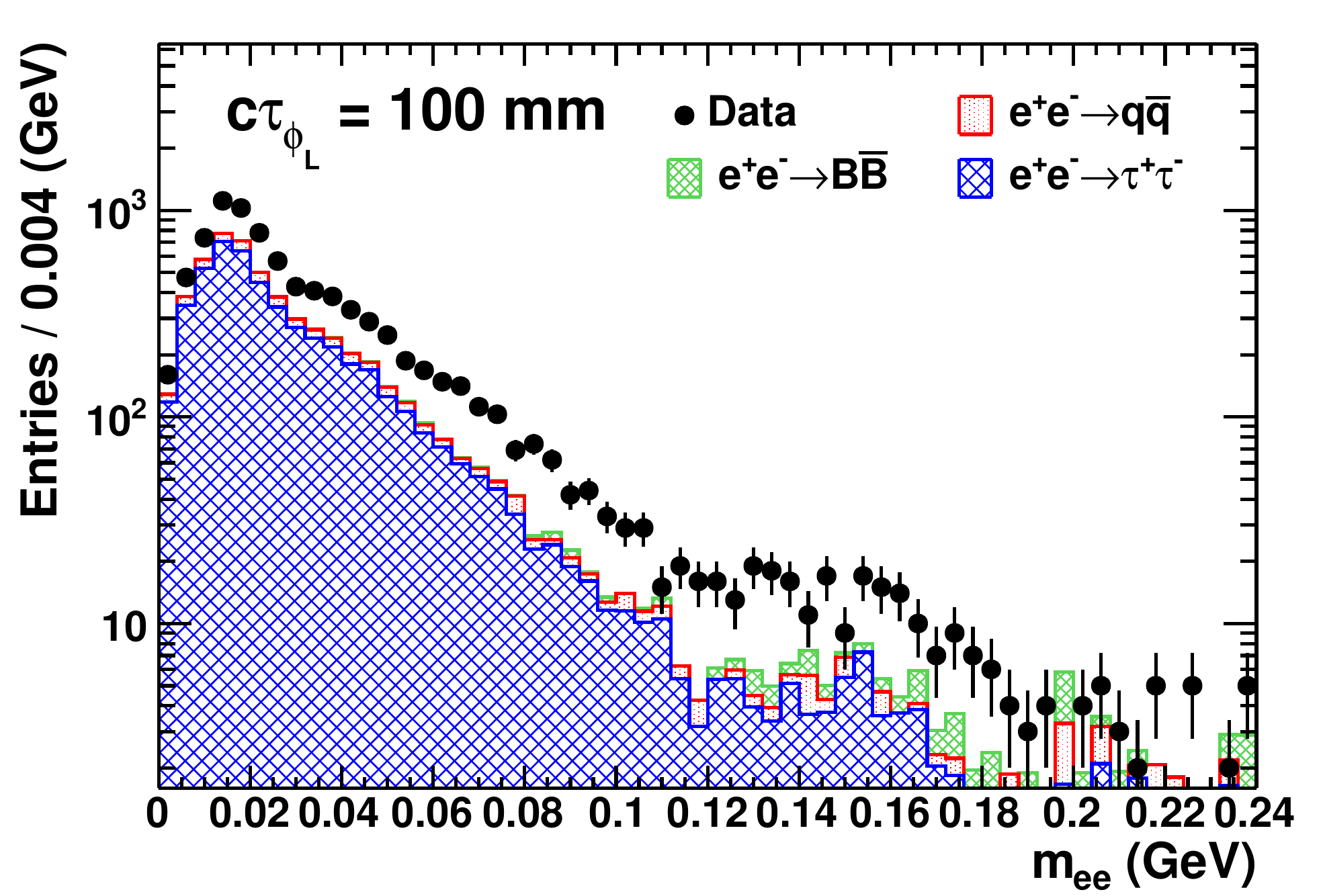}
\end{center}
\caption{The distribution of the dielectron invariant mass for the (top) $c\tau_{\phi_L}=1 \rm \, mm$, 
(middle) $c\tau_{\phi_L}=10  \rm \, mm$, and (bottom) $c\tau_{\phi_L}=100 \rm \, mm$ samples, together with 
simulated predictions for the indicated processes normalized to the integrated luminosity of the data (stacked histograms).}
\label{Fig2}
\end{figure}

The final selection of $\pL$ candidates for each lifetime selection and decay channel is made by applying a mass-dependent 
criterion on the corresponding BDT score that maximizes signal sensitivity. The distributions of the resulting dielectron 
and dimuon masses for prompt decays are shown in Fig.~\ref{Fig1}, and spectra for other lifetimes for $\pL \rightarrow \epem$ 
decays are shown in Fig.~\ref{Fig2}, together with the dominant background components among the set of simulated MC samples. 
The differences between the data and summed-MC distributions are mainly due to processes that are not simulated, dominated by 
ISR production of high-multiplicity QED and hadronic events as well as two-photon processes. Peaking contributions from 
$\jpsi \rightarrow \mpmm$ and $\psitwos \rightarrow \mpmm$ decays are also seen, and the corresponding regions are excluded 
from the signal search. In addition, the dielectron spectrum for $c\tau_{\phi_L}=1 \rm \,mm$ features a broad enhancement from $\pi^0\rightarrow\gamma\gamma$ decays in which one or both photons convert to $\epem$ pairs. Since this feature is much broader 
than the signal, we do not exclude this mass region but instead treat it as an additional background component. No statistically 
significant $\piz$ component is observed for other values of $c\tau_{\phi_L}$.

We extract the signal yield for the different lifetimes and final states separately by scanning the corresponding mass spectrum 
in steps of the signal mass resolution, $\sigma$. The latter is estimated by performing fits of a double-sided Crystal 
Ball function~\cite{CrystalBall} to each signal MC sample and interpolating the results to the full mass range. 
The resolution ranges from $1\MeV$ near $\mpL=40\MeV$ for $c\tau_{\phi_L} = 100 \rm \, mm$ to $50\MeV$ near $\mpL = 7.0 \GeV$ for 
prompt decays. The signal MC predictions are validated with samples of $\KS \rightarrow \pipm$ and 
$\psitwos \rightarrow \pipm \jpsi, \jpsi \rightarrow \mpmm$ decays; agreement with the data is observed. 
For each mass hypothesis, we perform an unbinned likelihood fit over an interval varying between $20-50\sigma$ (fixed 
to $60 \sigma$) for the dielectron (dimuon) final state. To facilitate the background description, the reduced dimuon 
mass, $m_R = (m^2_{\mu\mu} - 4m_\mu^2)^{1/2}$, is used for $2 m_\mu < m_{\phi_L} < 260 \MeV$. In that region, fits are 
performed over a fixed interval $m_R < 0.2 \GeV$.

The likelihood function includes contributions from signal, continuum background, and, where needed, peaking components 
describing the $\pi^0$, $\jpsi\rightarrow \mu^+\mu^-$, and $\psitwos \rightarrow \mu^+\mu^-$ resonances. The signal 
probability density function (pdf) is described by a non-parametric kernel density function modeled directly from the 
signal MC mass distribution. An algorithm based on the cumulative density function~\cite{Read:1999kh} is used to 
interpolate the pdf between simulated mass points. The uncertainty associated with this procedure is on average 4\% (3\%) 
of the corresponding statistical uncertainty for the dielectron (dimuon) analysis.

The dielectron continuum background is modeled by a second-order polynomial for the $c\tau_{\phi_L}=100 \rm \, mm$ sample and 
by a second-order polynomial plus an exponential function for the other lifetimes. The peaking $\pi^0$ shape for the 
$c\tau_{\phi_L}=1 \rm \, mm$ sample is determined from sideband data obtained by applying all selection criteria, but requiring 
the $\chi^2/n.d.f.$ of the kinematic fit to be greater than 3. The peaking $\piz$ yield and all the continuum background 
parameters are determined in the fit. To assess systematic uncertainties, we repeat the fits with a third-order 
polynomial for the continuum background, vary the width of the $\pi^0$ shape within its uncertainty, or include 
a $\piz$ component for all lifetime samples. The resulting systematic uncertainties are typically at the level 
of the statistical uncertainty, but dominate the total uncertainty for several mass hypotheses in the vicinity of the $\piz$ peak.

The reduced mass distribution of the dimuon continuum background is modeled by a third-order polynomial constrained to 
intersect the origin, and the dimuon continuum is described by a second-order polynomial at higher masses. The shape 
of the $\jpsi$ and $\psitwos$ resonances are fixed to the predictions of the corresponding MC samples, but their yields 
cannot be accurately estimated from MC simulations and are therefore left to float freely in the fit. A range of $\pm 50 \MeV$ 
around the nominal $\jpsi$ and $\psitwos$ masses is therefore excluded from the search. The systematic uncertainty associated 
with the choice of the background model, assessed by repeating the fits with alternative descriptions, is typically at the 
level of a few events, but can be as large as half the statistical uncertainty for a few points in the high mass region, where 
statistical precision is limited.

The fitted signal yields and statistical significances are presented in the Supplemental Material~\cite{SPM}, together with 
a few examples of fits. The bias in the fitted values is determined from pseudo-experiments to be negligible compared to 
the statistical uncertainties. Since the systematic uncertainty associated with the choice of background model can be 
large in the dielectron channel, we define the signal significance as the smallest of the significance values determined 
from each background model. Including trial factors, the largest significance is $1.4\sigma$ observed near $\mpL = 2.14 \GeV$, 
consistent with the null hypothesis.

The signal efficiency varies between $0.2\%$ for $\mpL=40 \MeV$ and $c\tau_{\phi_L}=100 \rm \, mm$, to $26\%$ around $\mpL = 5\GeV$ 
for prompt decays. The effect of ISR, not included in the samples generated by \textsc{MadGraph}, is assessed by simulating events 
with \textsc{Pythia} 8 using the matrix elements calculated by \textsc{MadGraph}, and reweighting this sample to match 
the $p_{\rm T}$ distribution of the $\pL$ predicted by \textsc{MadGraph}. The resulting change in efficiency is found 
to be about 4\% over the full mass range covered by the dielectron channel, and varies from 7\% near the dimuon 
threshold to less than 1\% at $\mpL \sim 7 \GeV$. Half the value of each of these differences is propagated as a systematic
uncertainty  in the signal yield. A correction factor of 0.98 (0.93) on the signal efficiency is included 
for the dielectron (dimuon) final state to account for differences between data and simulation in track and neutral 
reconstruction efficiencies, charged particle identification, and trigger efficiencies. The correction for the dielectron 
channel is derived from a sample of $\KS\rightarrow \pi^+\pi^-$ produced in $\tau$ decays, while that for the dimuon channel 
is assessed from the BDT score distribution for events in which the missing transverse momentum is greater than 
$2 \GeV$, a region where the contribution of unsimulated background components can be neglected. An uncertainty of 
3.8\% (4.0\%) in the dielectron (dimuon) efficiency correction is propagated as a systematic uncertainty. 

\begin{figure}[tb]
\begin{center}
  \includegraphics[width=0.48\textwidth]{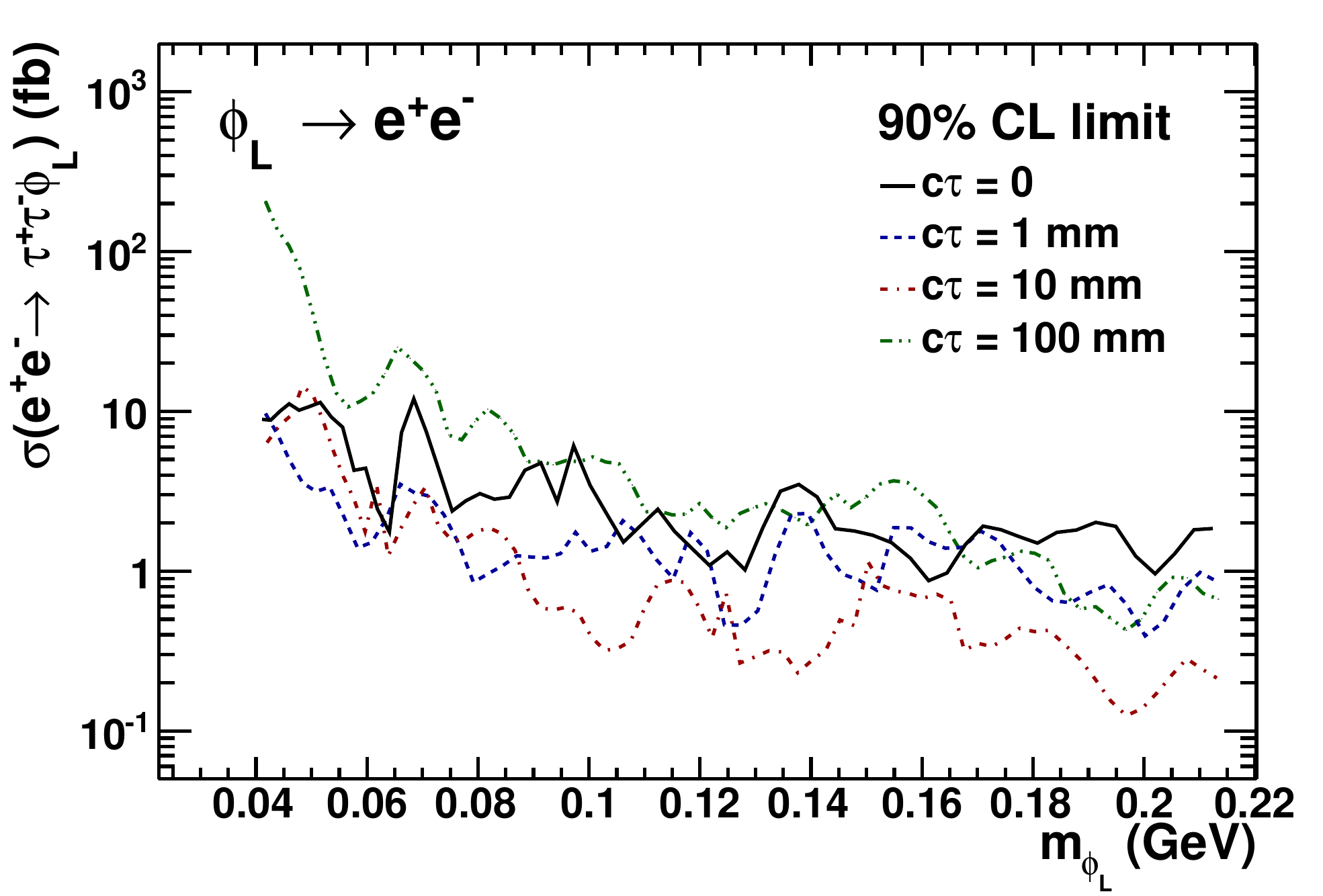}
  \includegraphics[width=0.48\textwidth]{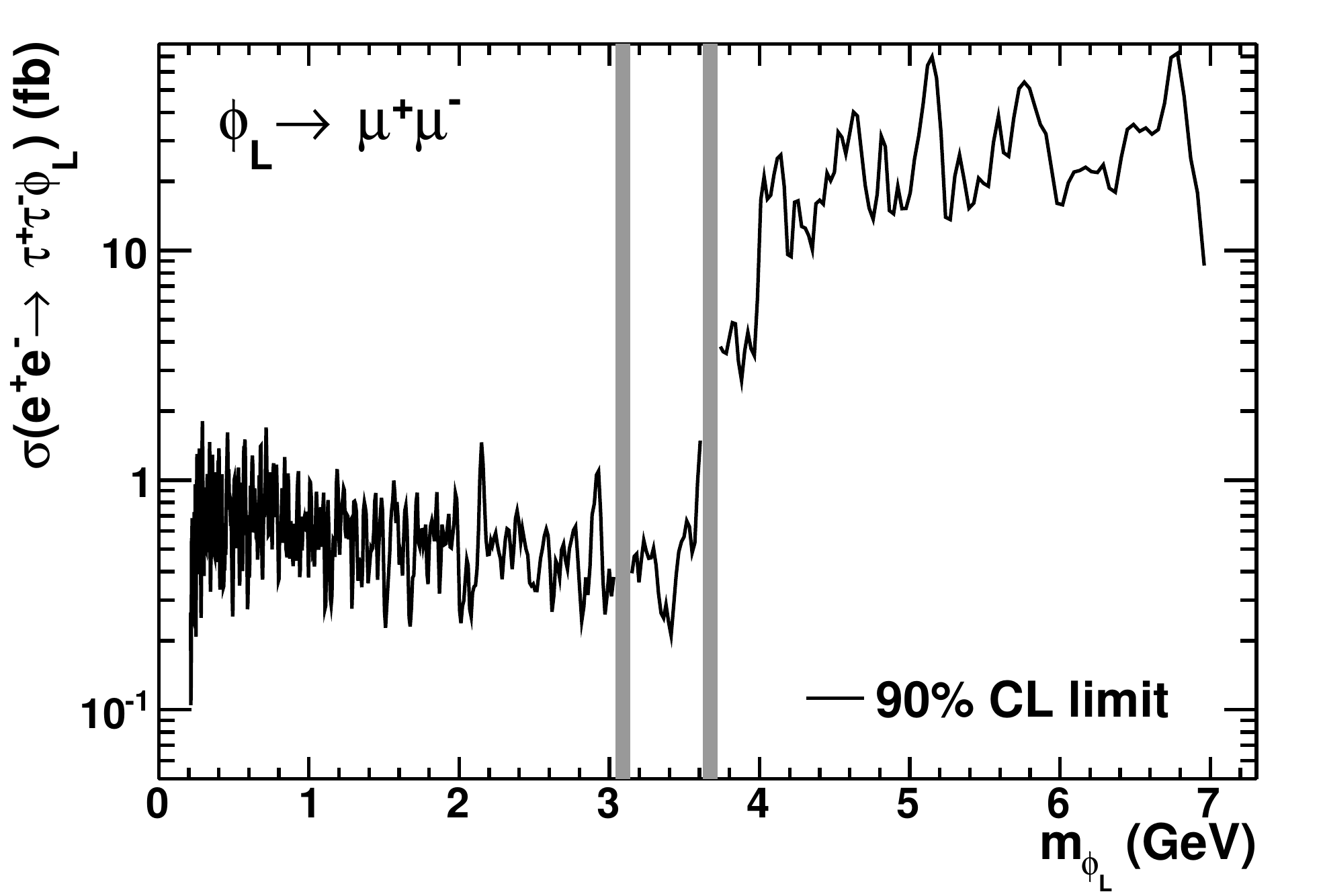}
\end{center}
\caption{The 90\% CL upper limits on the $\sigma(e^+e^- \rightarrow \tptm \pL)$ cross section at the $\Y4S$ resonance 
derived from (top) the dielectron and (bottom) dimuon final states. The gray bands indicate the regions excluded from 
the search around the nominal $\jpsi$ and $\psitwos$ masses.}
\label{Fig3}
\end{figure}

The $e^+e^- \rightarrow \tptm \pL$ cross section at the $\Upsilon(4S)$ energy is derived for each lifetime and 
final state by taking into account the variation of the cross section and signal efficiencies with the beam energy 
and the $\pL\rightarrow \ell^+\ell^-$ branching fraction:
$$ \sigma_{4S} = \frac{N_{sig}} {\sum\limits_{i=2S,3S,4S}(\frac{\sigma_{th,i}}{\sigma_{th,4S}} \epsilon_{i}{\cal L}_{i}) 
\,\, BF(\pL \rightarrow \ell^+\ell^-)},$$
where $N_{sig}$ denotes the number of signal events, and $\sigma_{th,nS}$, $\epsilon_{nS}$ and ${\cal L}_{nS}$ ($n=2,3,4$) 
are the theoretical $e^+e^- \rightarrow \tptm \pL$ cross section, signal efficiency, and data luminosity at the $\Upsilon(nS)$ 
center-of-mass energy, respectively. In the absence of a significant signal, Bayesian upper limits at 90\% confidence 
level (CL) on the cross sections are derived by assuming a uniform prior in the cross section. Systematic effects are 
taken into account by convolving the likelihood with a Gaussian having a width equal to the systematic uncertainty. The 
uncertainties due to the luminosity (0.6\%)~\cite{Lees:2013rw} and the limited statistical precision of the signal MC 
sample (1--4\%) are incorporated.  The resulting limits are shown in Fig.~\ref{Fig3}. The sharp increase just above the 
ditau threshold is a reflection of the $\pL \rightarrow \mpmm$ branching fraction decreasing quickly in favor of the 
$\tptm$ final state. The limit on the production cross section of a scalar $S$ without any assumptions on other 
decay modes is presented in the Supplemental Material~\cite{SPM}.

\begin{figure}[htb]
\begin{center}
  \includegraphics[width=0.48\textwidth]{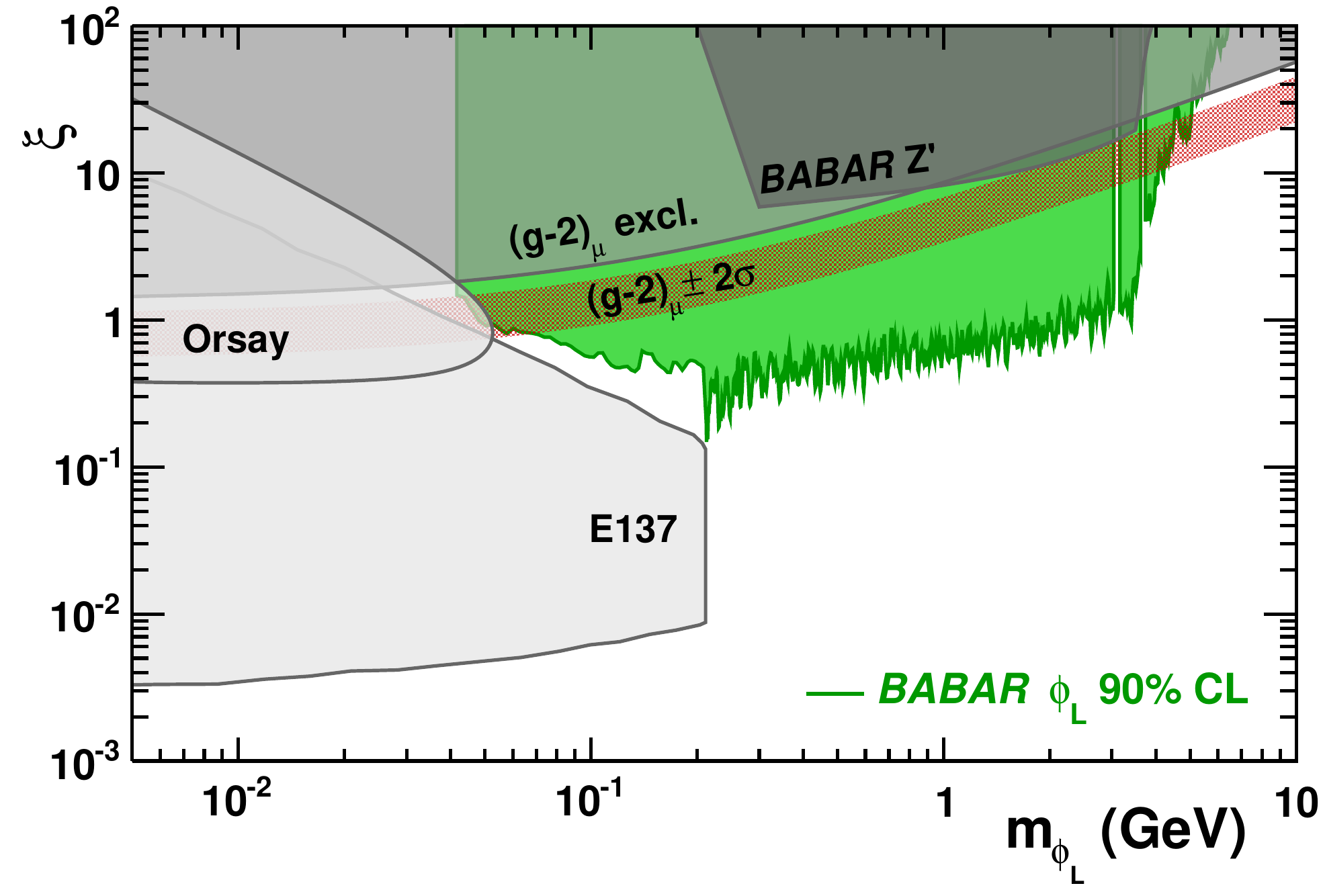}
\end{center}
\caption{The 90\% CL limits on the coupling $\xi$ as a function of the $\pL$ mass (green shaded area), together 
with existing constraints~\protect{\cite{TheBABAR:2016rlg,Bjorken:1988as,Davier:1989wz,Liu:2016qwd,Liu:2020qgx}} (gray 
shaded areas) and the parameter space preferred by the muon anomalous magnetic 
moment~\protect{\cite{Batell:2016ove,Liu:2020qgx}} (red band).}
\label{Fig4}
\end{figure}

The limits on the scalar coupling $\xi$, presented in Fig.~\ref{Fig4}, are derived with an iterative procedure that 
accounts for a potentially long $\pL$ lifetime. An estimate of $\xi$ is first chosen, and the corresponding lifetime 
and cross section are calculated~\cite{Batell:2016ove}. These values are compared to the cross section limit interpolated at that lifetime, 
and the estimate of the coupling is updated. The procedure is iterated until convergence is obtained. Bounds at the level of $0.5-1$ are 
set on the dielectron final state, corresponding to $c\tau_{\phi_L}$ values of the order of 1\,cm, and limits down to 0.2 are 
derived for dimuon decays. These results are approximately an order of magnitude smaller than the couplings favored by the muon 
anomalous magnetic moment below the ditau threshold~\cite{Batell:2016ove} and rule out a substantial fraction of previously 
unexplored parameter space at 90\% CL.

In summary, we report the first model-independent search for the direct production of a new dark leptophilic scalar. The 
limits significantly improve upon the previous constraints over a large range of masses, almost entirely ruling out the 
remaining region of parameter space below the dimuon threshold. More significantly, this search excludes the possibility of 
the dark leptophilic scalar accounting for the observed discrepancy in the muon magnetic moment for almost all $\pL$ masses 
below $4\GeV$. Since these results rely only on $\pL$ production in association with tau leptons and its subsequent 
leptonic decay, they can also be reinterpreted to provide powerful constraints on other leptonically decaying new 
bosons interacting with tau leptons. The Belle II experiment should be able to further probe these possibilities, and 
cover the remaining parameter space above the beam dump constraints. 

We are grateful for the 
extraordinary contributions of our \pep2\ colleagues in
achieving the excellent luminosity and machine conditions
that have made this work possible.
The success of this project also relies critically on the 
expertise and dedication of the computing organizations that 
support \babar.
The collaborating institutions wish to thank 
SLAC for its support and the kind hospitality extended to them. 
This work is supported by the
US Department of Energy
and National Science Foundation, the
Natural Sciences and Engineering Research Council (Canada),
the Commissariat \`a l'Energie Atomique and
Institut National de Physique Nucl\'eaire et de Physique des Particules
(France), the
Bundesministerium f\"ur Bildung und Forschung and
Deutsche Forschungsgemeinschaft
(Germany), the
Istituto Nazionale di Fisica Nucleare (Italy),
the Foundation for Fundamental Research on Matter (The Netherlands),
the Research Council of Norway, the
Ministry of Education and Science of the Russian Federation, 
Ministerio de Econom\'{\i}a y Competitividad (Spain), the
Science and Technology Facilities Council (United Kingdom),
and the Binational Science Foundation (U.S.-Israel).
Individuals have received support from 
the Marie-Curie IEF program (European Union) and the A. P. Sloan Foundation (USA). 


\clearpage
\onecolumngrid
\begin{center} {\bf \large Supplemental Material for BABAR-PUB-20/003} \end{center}
\begin{center} {\it \large Search for a Dark Leptophilic Scalar in $e^+e^-$ Collisions} \end{center}
\vspace{0.5 cm}

Additional details and figures for the dark leptophilic scalar search are presented in this Supplemental Material.

\begin{table}[h]
\begin{center}
\caption{List of variables used as input to the dimuon boosted decision trees.}
\begin{tabular}{|l|}
\hline
  Ratio of second to zeroth Fox-Wolfram moment of all tracks and neutrals. \\ 
  Invariant mass of the four track system, assuming the pion (muon) mass for the tracks originating from the tau ($\pL$) decays.\\
  Invariant mass and transverse momentum of all tracks and neutrals.\\
  Invariant mass squared of the system recoiling against all tracks and neutrals.\\
  Transverse momentum of the system recoiling against all tracks and neutrals.\\
  Number of neutral candidates with an energy greater than $50\MeV$.\\
  Invariant masses of the three track systems formed by the $\pL$ and the remaining positively or negatively charged tracks.\\
  Momentum of each track from $\pL$ decays.\\
  Angle between the two tracks produced by the tau decay.\\
  Variable indicating if a track has been identified as a muon or an electron by PID algorithm for each track.\\
\hline
\end{tabular}
\end{center}
\end{table}

\begin{table}[h]
\begin{center}
\caption{List of variables used as input to the dielectron boosted decision trees.}
\begin{tabular}{|l|}
\hline
  Transverse momentum of the system recoiling against all tracks and neutrals.\\
  Energy of the system recoiling against all tracks and neutrals.\\
  Number of tracks identified as electron candidates by a PID algorithm applied to each track.\\
  Angle between $\pL$ candidate momentum and closest track produced in tau decay.\\
  Angle between $\pL$ candidate momentum and farthest track produced in tau decay.\\
  Angle of $\pL$ candidate relative to the beam in the center-of-mass frame.\\
  Angle between the two tracks produced by the tau decay.\\
  Angle between $\pL$ candidate and nearest neutral candidate with $E>50 \MeV$. \\
  Energy of nearest neutral candidate (with $E>50\MeV$) to $\pL$ candidate.\\
  Total energy in neutral candidates, each of which has an energy greater than $50\MeV$.\\
  Distance between beamspot and $\pL$ candidate vertex.\\
  Uncertainty in the distance between beamspot and $\pL$ candidate decay vertex.\\
  $\pL$ candidate vertex significance, defined by the  beamspot-vertex distance divided by its uncertainty.\\
  Angle between the $\pL$ candidate momentum, and line from beamspot to $\pL$ decay vertex.\\
  Distance of closest approach to beamspot of $e^-$ in $\pL$ candidate.\\
  Distance of closest approach to beamspot of $e^+$ in $\pL$ candidate.\\
  Transverse distance between $\pL$ decay vertex and best-fit common origin of $\tau$ candidates and $\pL$ candidate.\\
  $\chi^2$ of the kinematic fit to the $\pL$ and $\tau$ candidates constraining their origin to the same production point.\\
  $\chi^2$ of the kinematic fit of the $\pL$ candidate with the constraint that the $\epem$ pair is produced from a photon \\
  conversion in detector material.\\
  Dielectron mass for $\pL$ candidate when re-fit with the photon conversion constraint.\\
\hline
\end{tabular}
\end{center}
\end{table}

\begin{figure}[htb]
\begin{center}
  \includegraphics[width=0.45\textwidth]{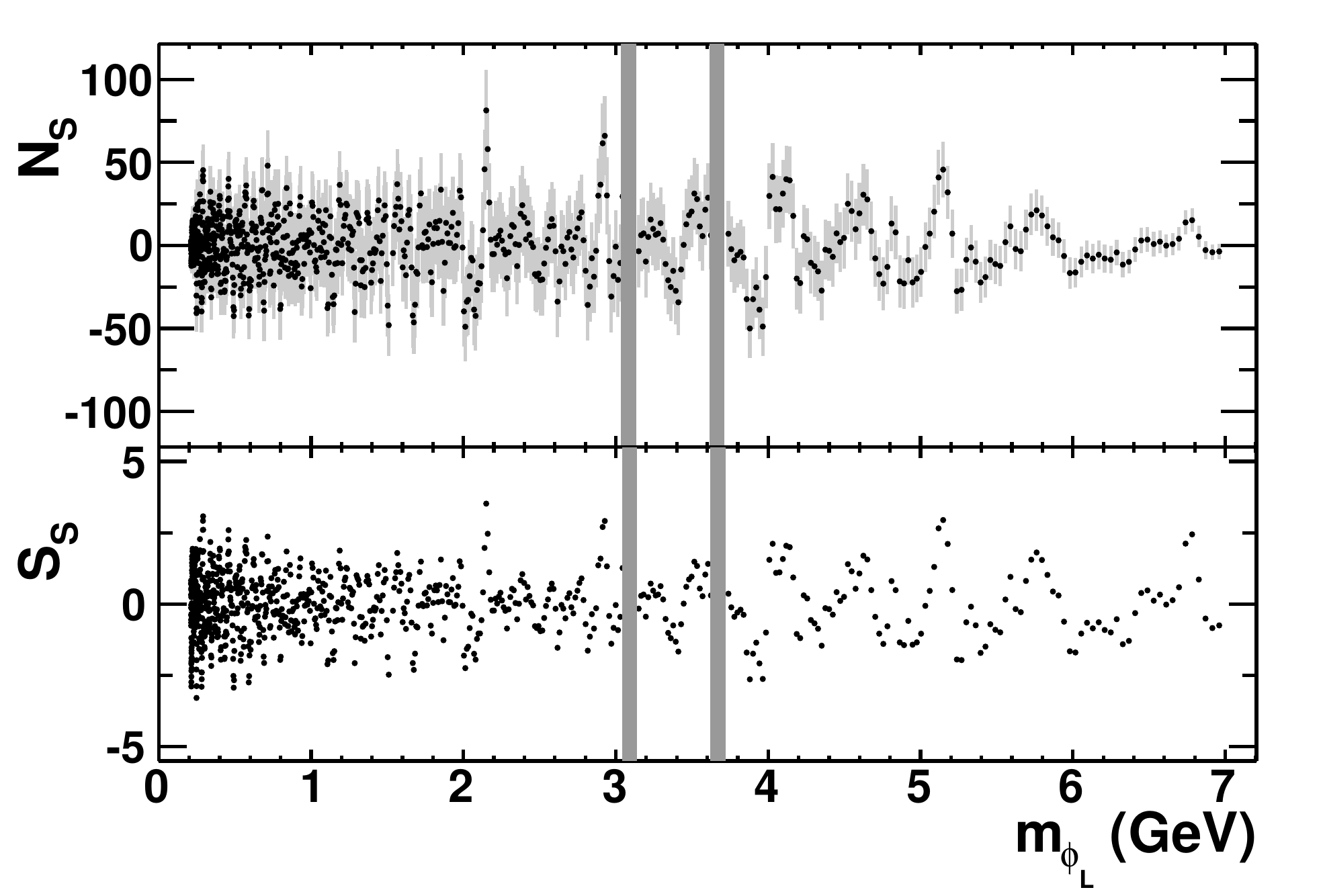}
  \includegraphics[width=0.45\textwidth]{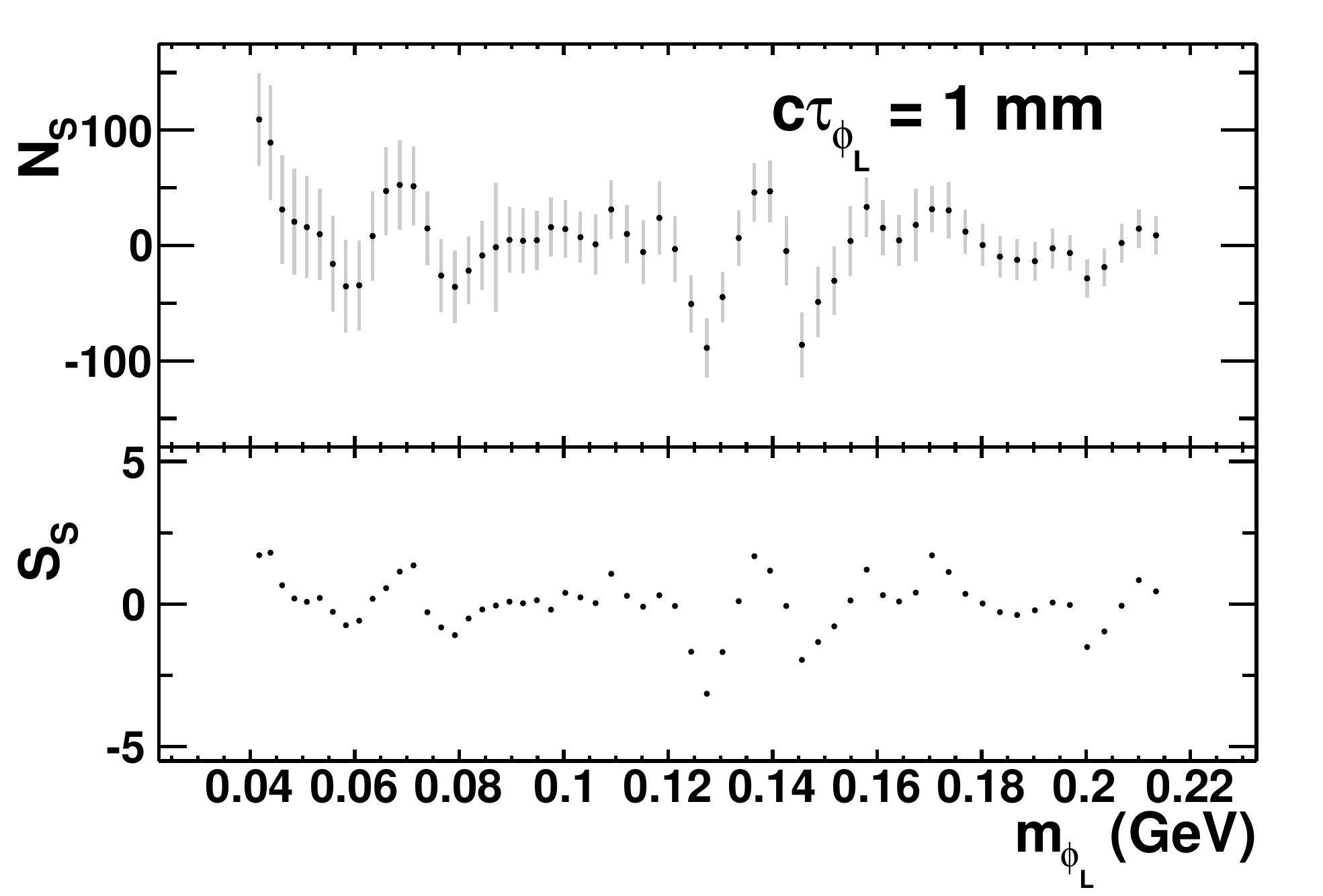}\\
  \includegraphics[width=0.45\textwidth]{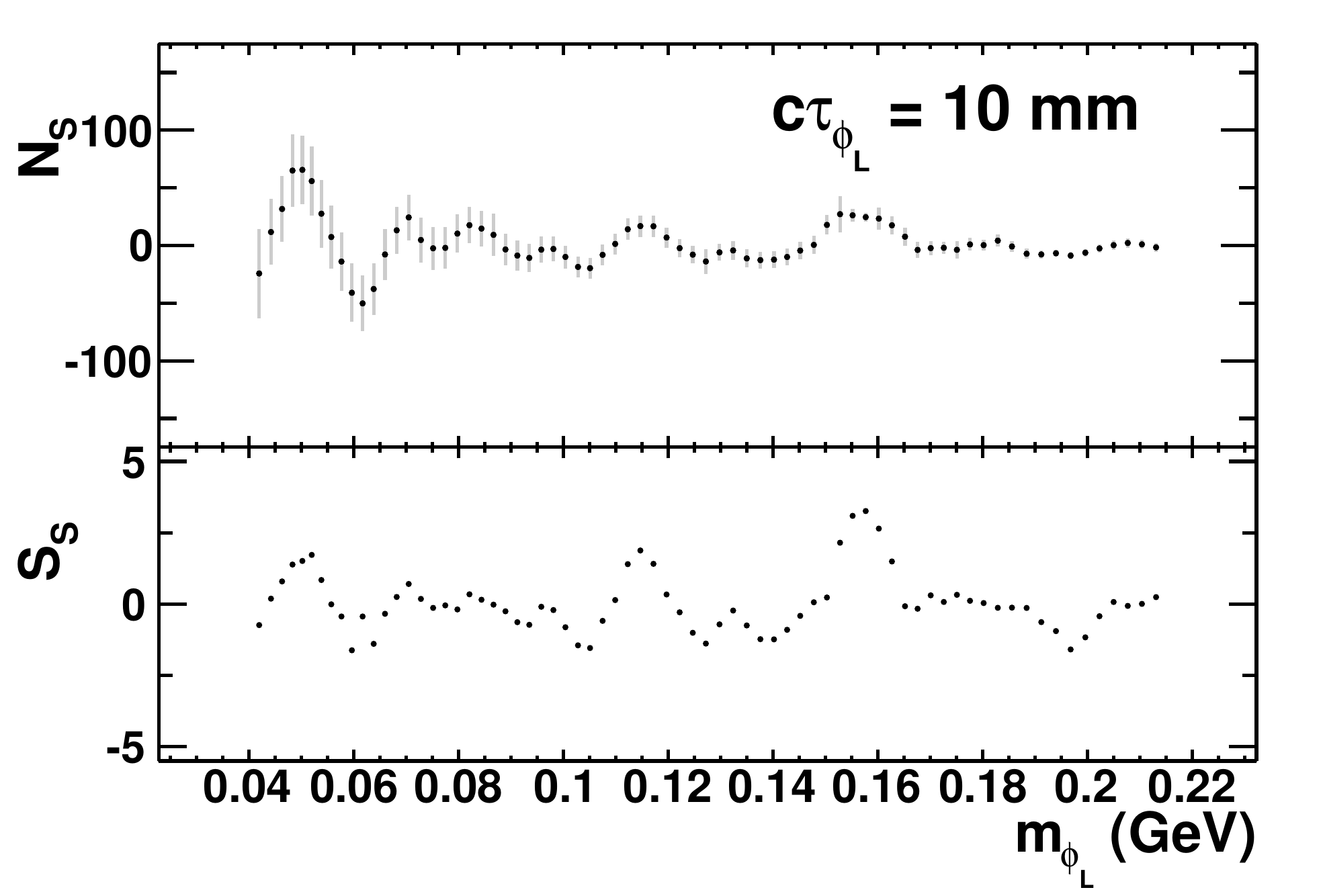}
  \includegraphics[width=0.45\textwidth]{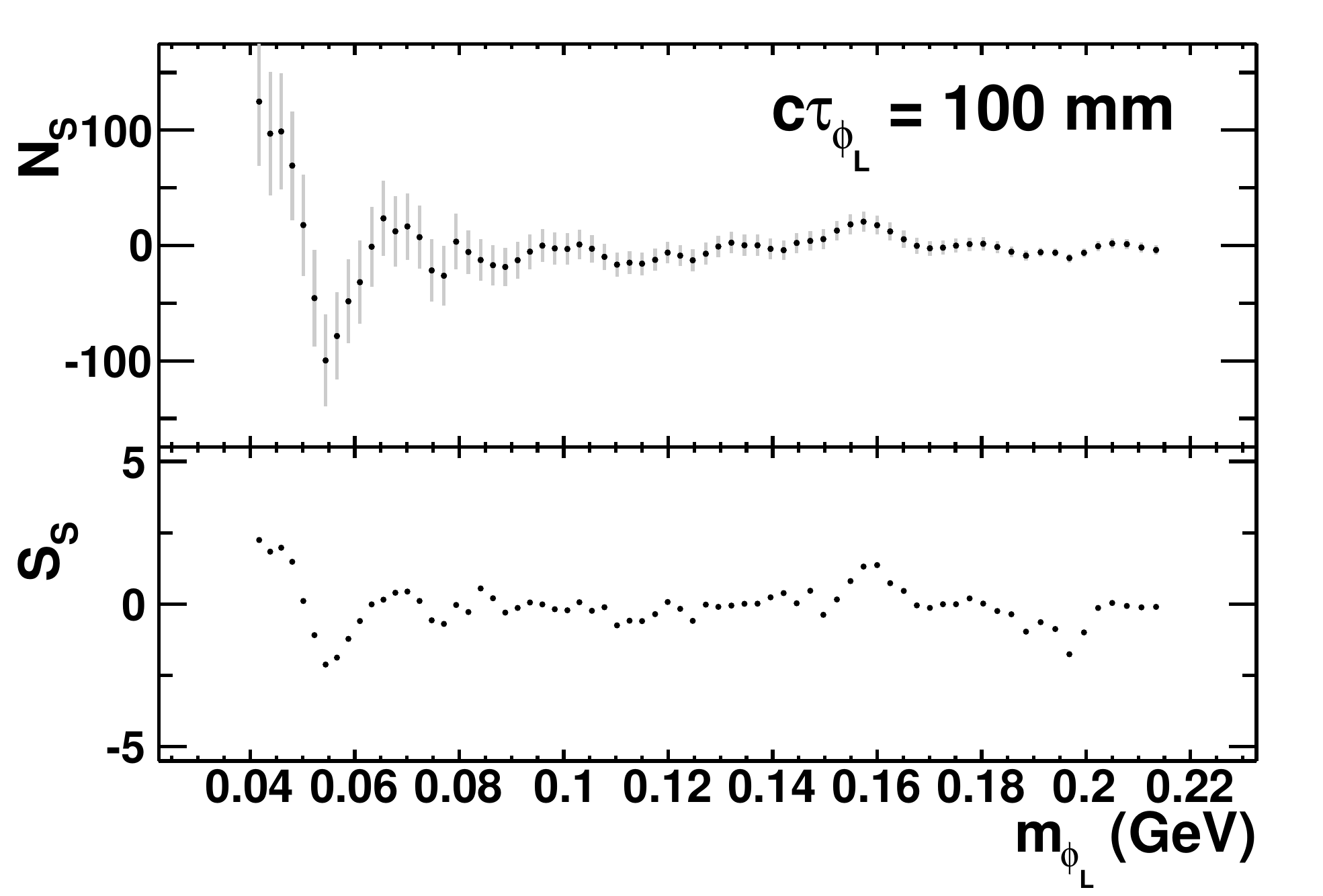}

\end{center}
\caption{The distribution of signal events ($N_s$) and local signal significance ($S_s$) from the fits 
as a function of the $\pL$ mass for (top left) prompt decays; (top right) $c\tau_{\phi_L}=1 \rm \, mm$; (bottom left) 
$c\tau_{\phi_L}=10 \rm \, mm$; (bottom right) $c\tau_{\phi_L}=100 \rm \, mm$. The prompt decays include 
contributions from both the dielectron and dimuon final states.}
\label{Fig6}
\end{figure}

\begin{figure}[htb]
\begin{center}
  \includegraphics[width=0.45\textwidth]{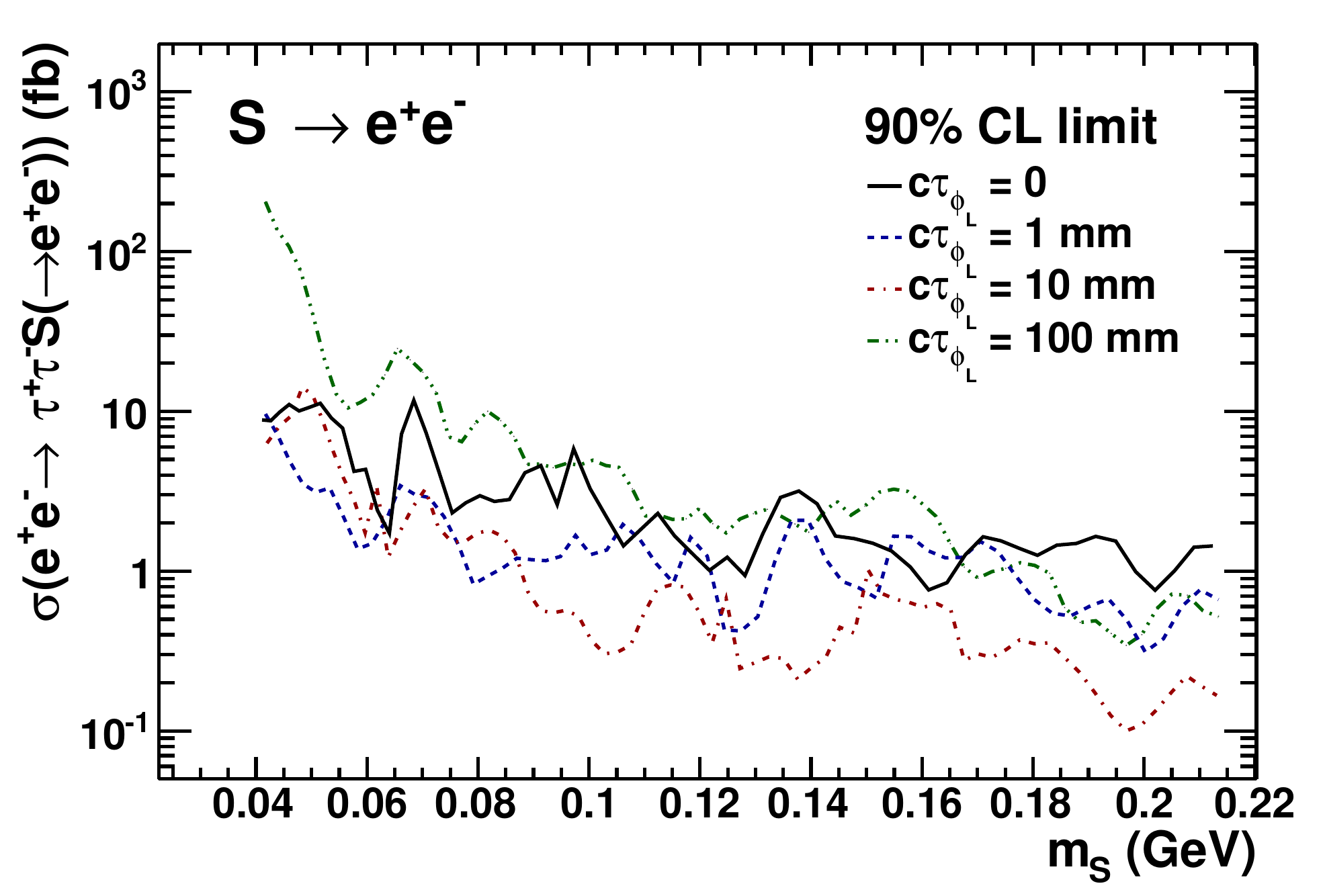}
  \includegraphics[width=0.45\textwidth]{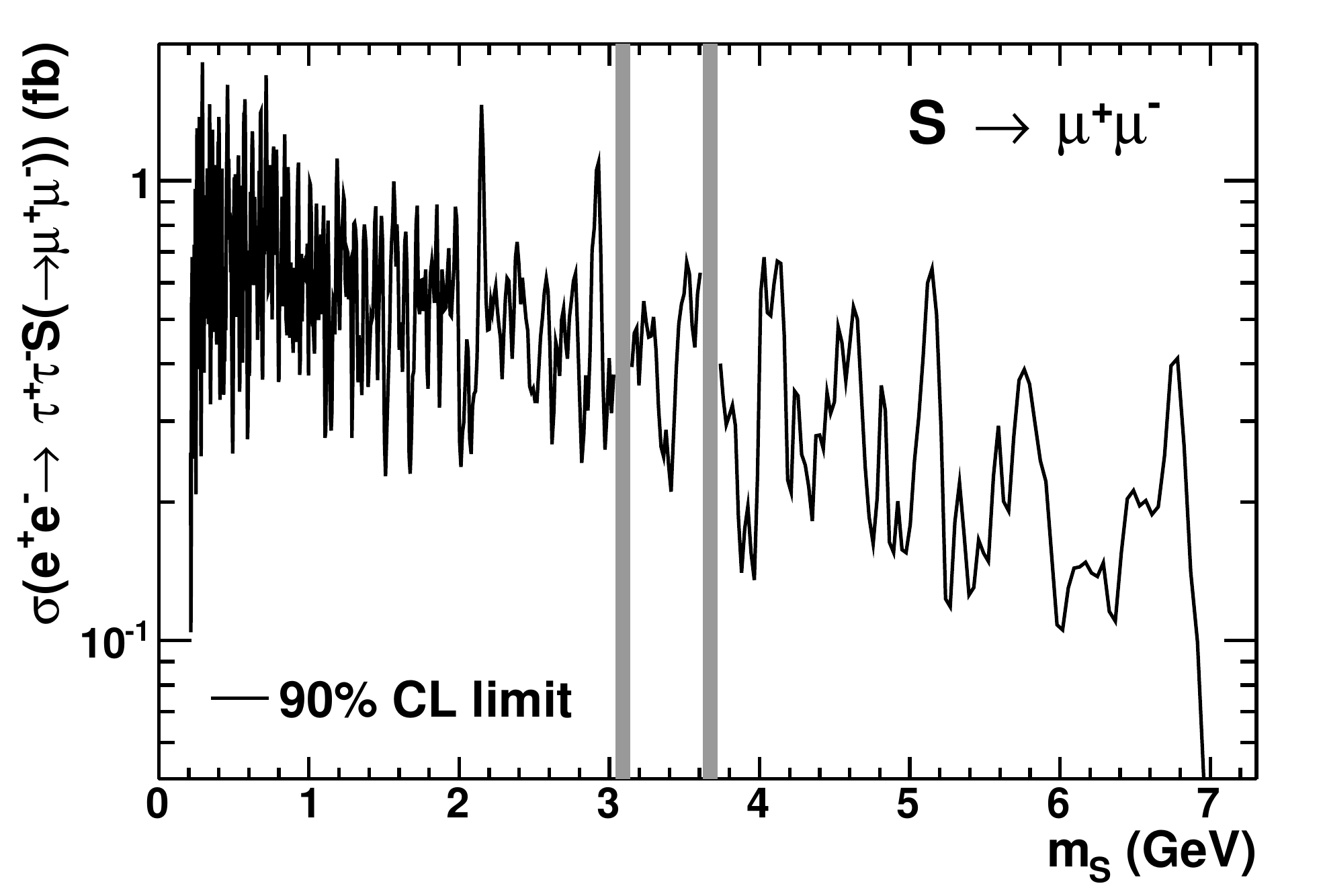}
\end{center}
\caption{The 90\% CL limits on (left) the $\sigma(e^+e^- \rightarrow \tptm S(S\rightarrow e^+e^-))$ 
and (right) the $\sigma(e^+e^- \rightarrow \tptm S(S\rightarrow \mu^+\mu^-))$ cross sections for the 
production of a generic scalar $S$ at the $\Y4S$ resonance. The gray bands indicate the regions excluded from 
the search around the nominal $\jpsi$ and $\psitwos$ masses.}
\label{Fig7}
\end{figure}

\begin{figure}[htb]
\begin{center}
\includegraphics[width=0.45\textwidth]{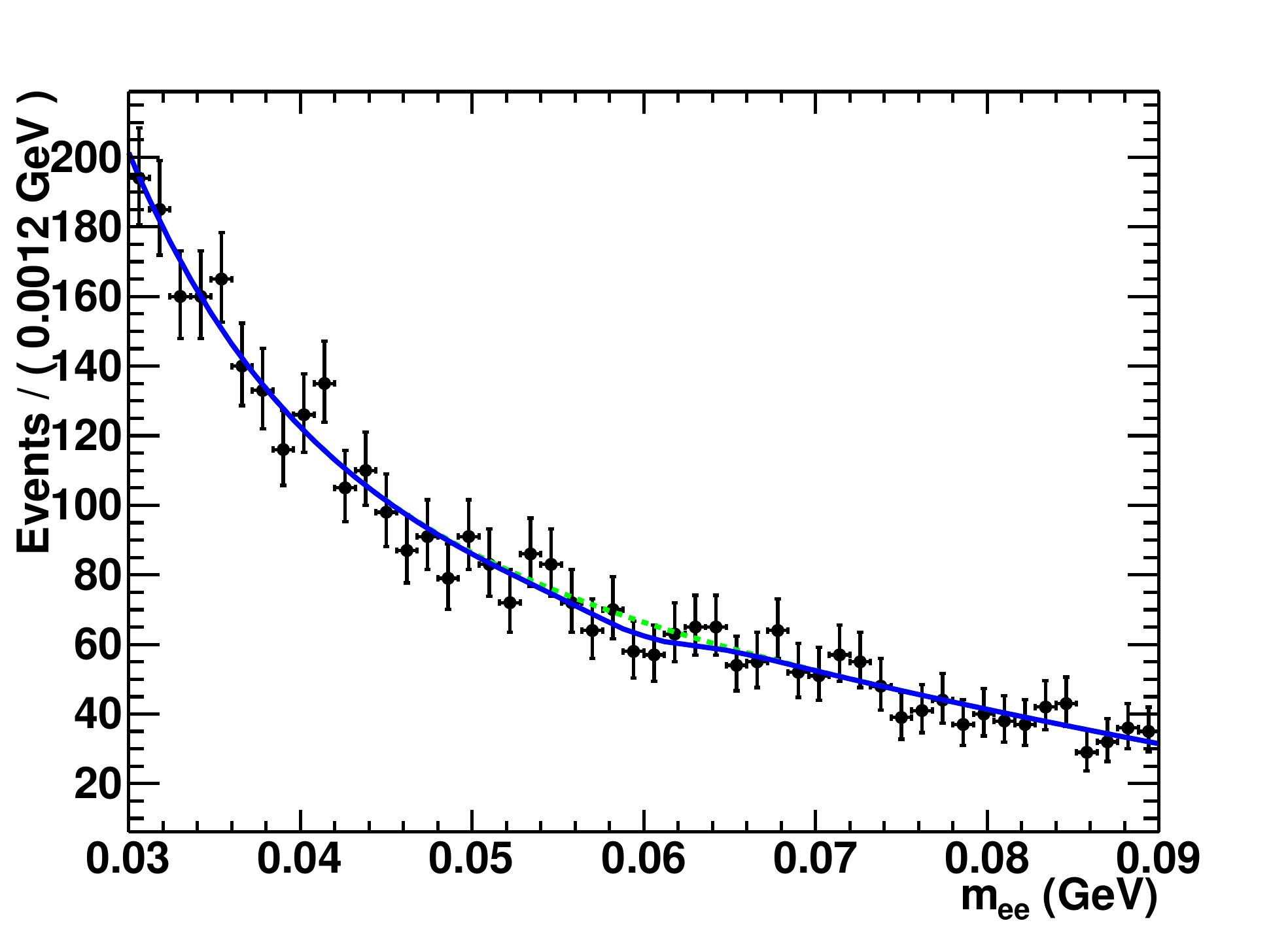} 
\includegraphics[width=0.45\textwidth]{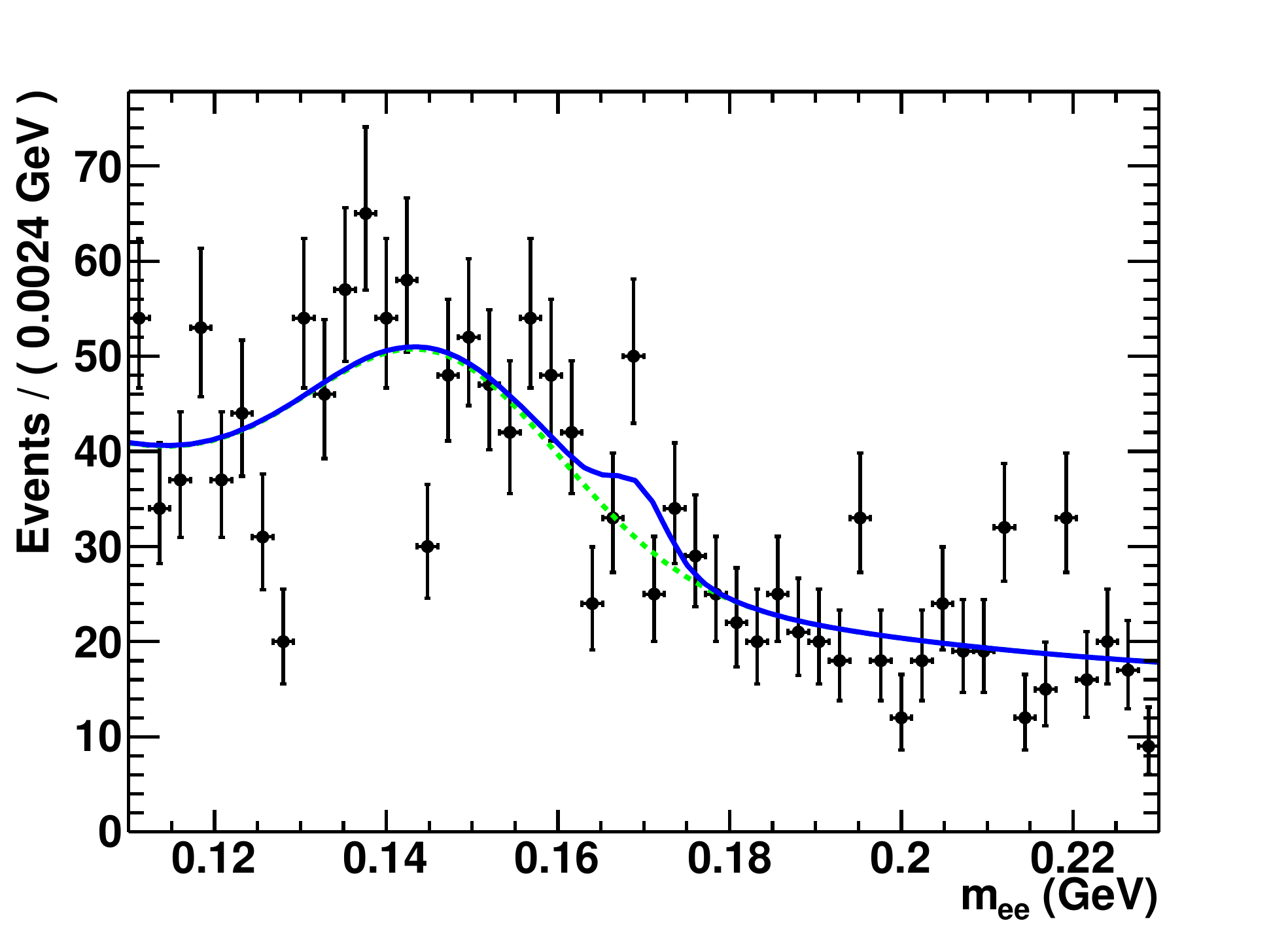} \\
\includegraphics[width=0.45\textwidth]{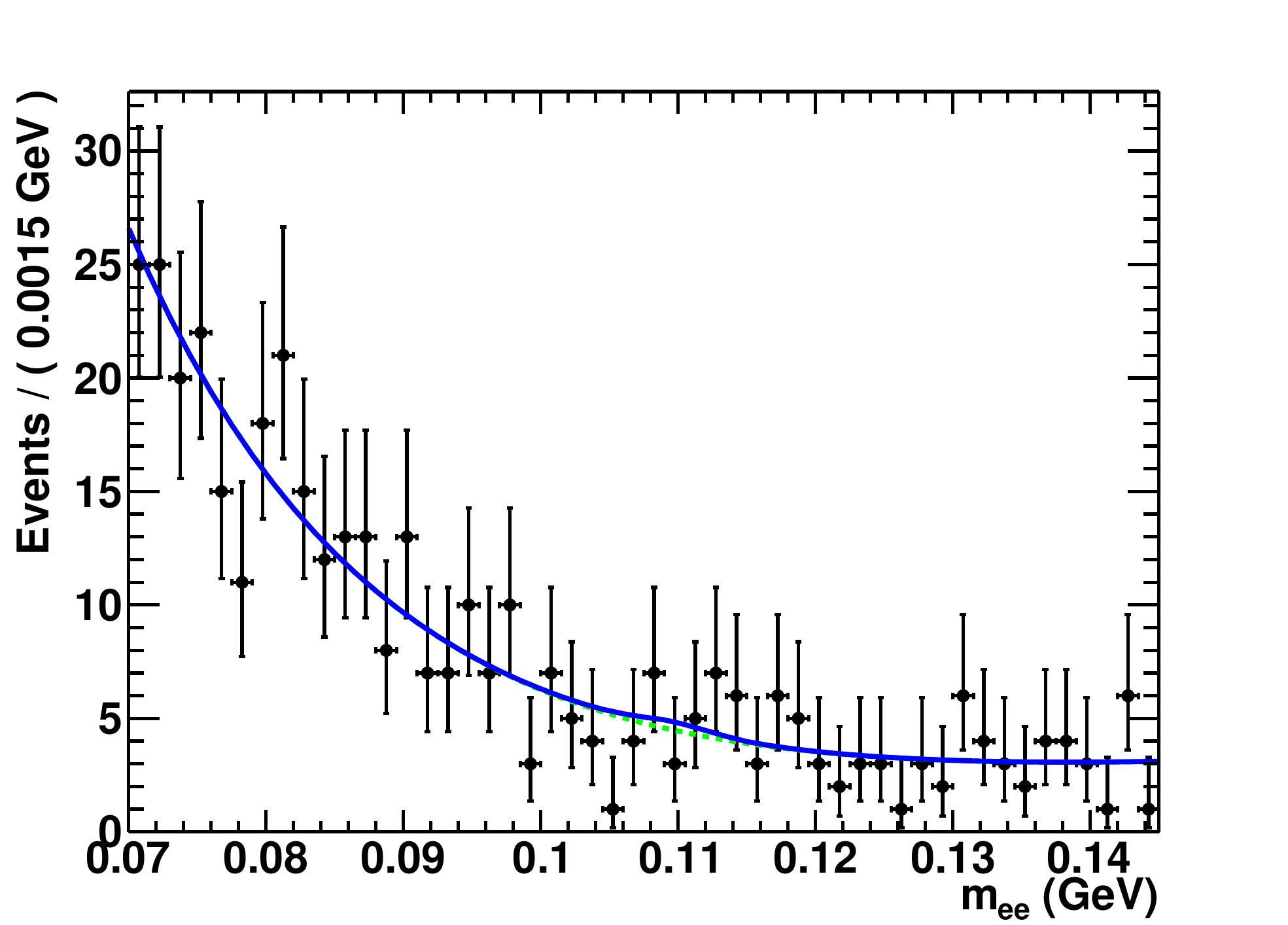} 
\includegraphics[width=0.45\textwidth]{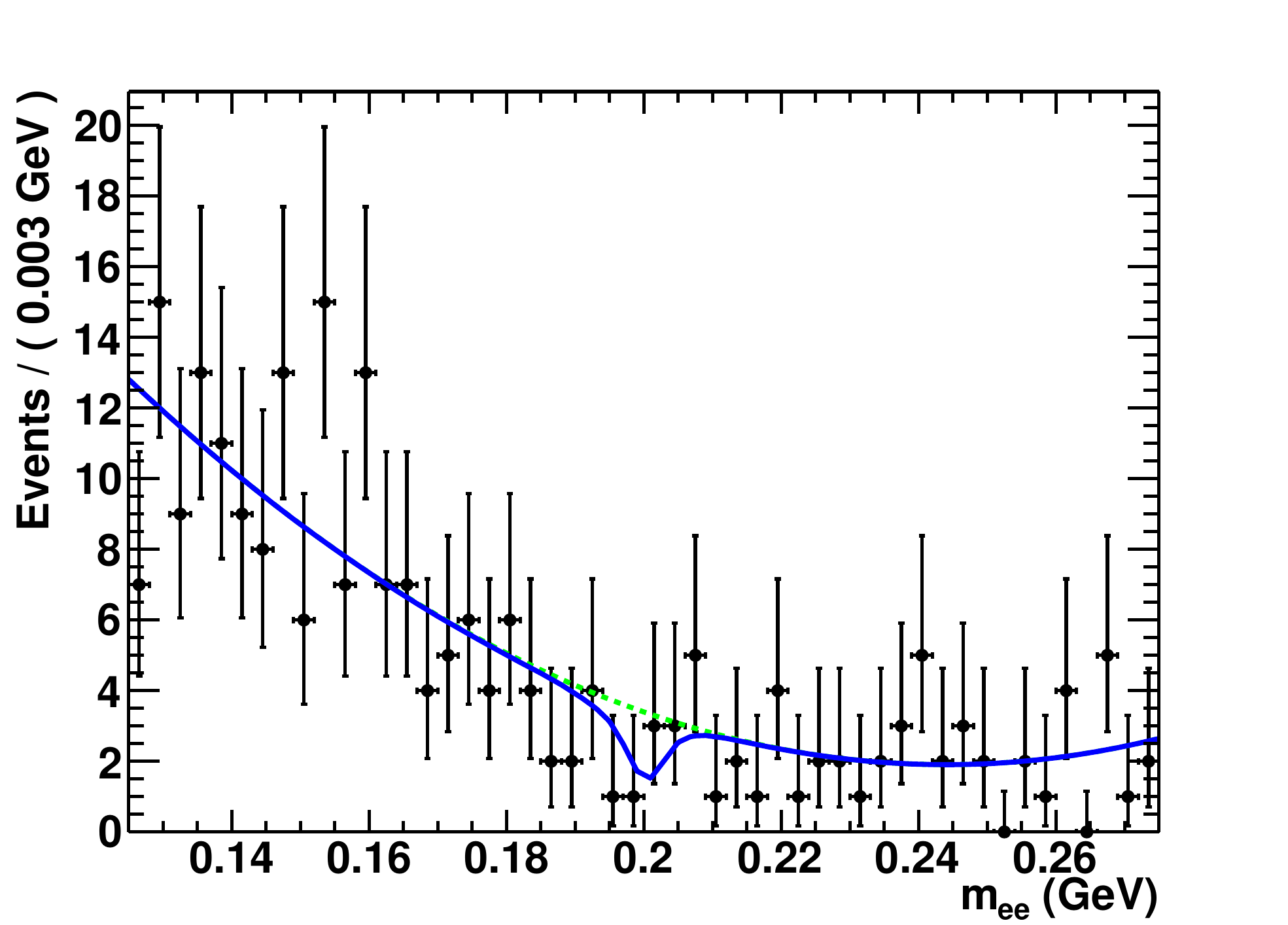} \\
\includegraphics[width=0.45\textwidth]{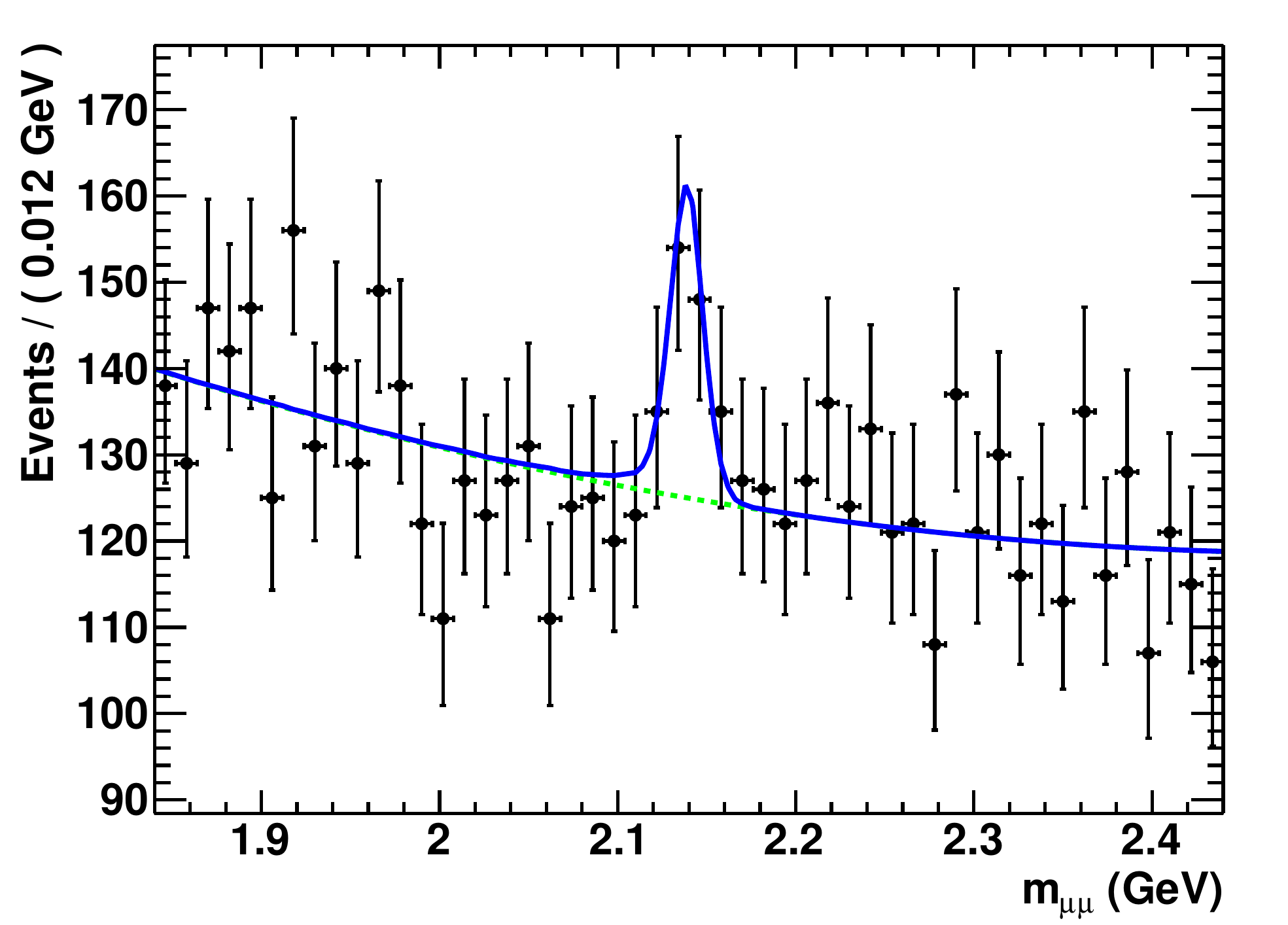} 
\includegraphics[width=0.45\textwidth]{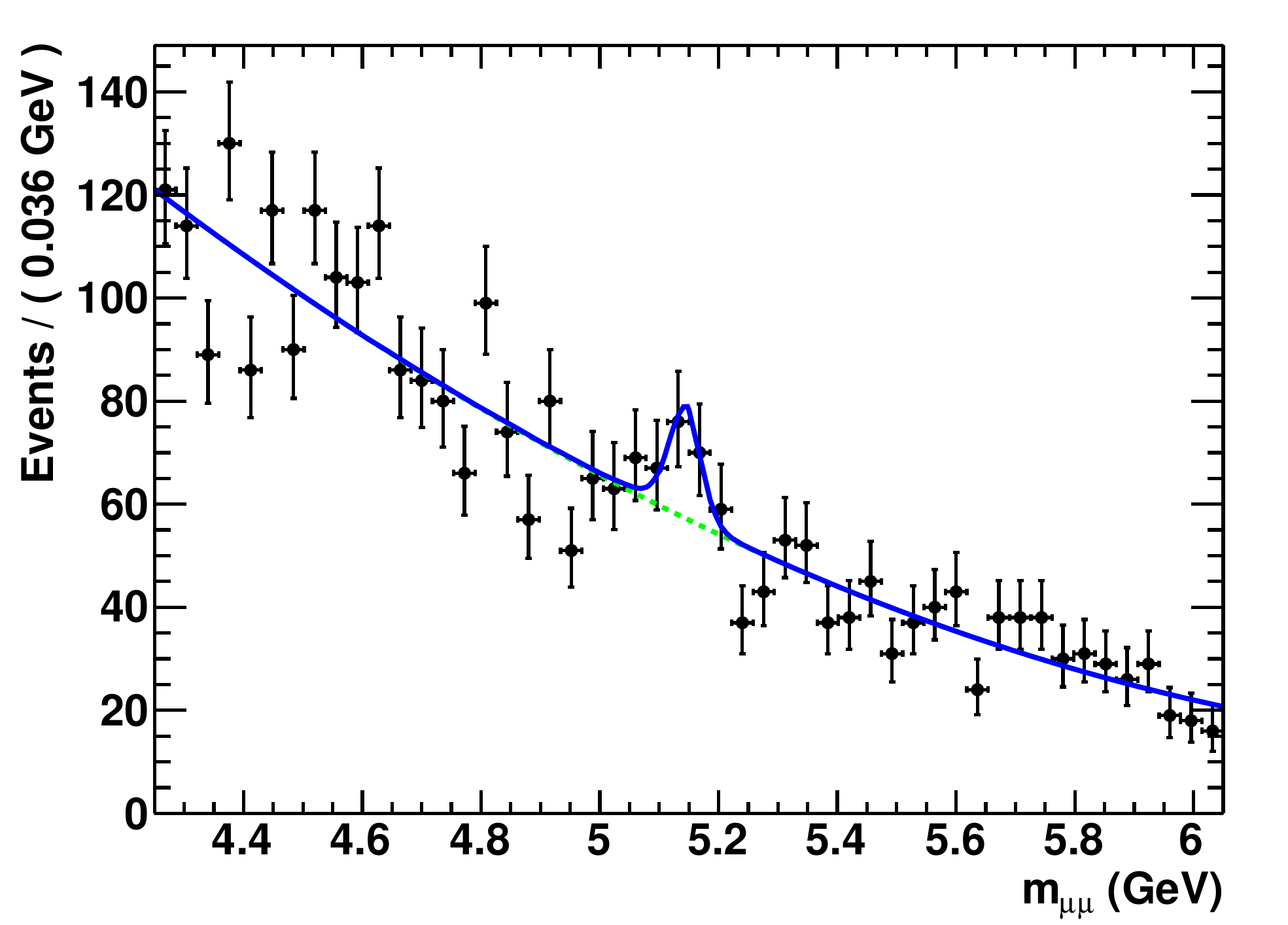} 
\end{center}
\caption
{Example of fits to (top) the dielectron mass with $c\tau=1$ mm; (middle left)  $c\tau=10$ mm; (middle right) $c\tau=100$ mm;  (bottom) the dimuon mass. The full fit is shown as a solid blue line, and the background as a dashed green line.}
\label{Fig8}
\end{figure}


\begin{thebibliography}{9}
\bibitem{naturalUnits} Natural units ($\hbar=c=1$) are used throughout this paper.
\end{thebibliography}
\end{document}